\documentclass[aps,prb,twocolumn,showpacs,showkeys,superscriptaddress]{revtex4}
\usepackage{graphicx}
\usepackage{color}
\usepackage{tabularx}

\begin{document}

\title{X-ray magnetic circular dichroism in (Ge,Mn) compounds: experiments and modeling}

\author{Samuel~Tardif}
\altaffiliation{current affiliation: RIKEN, SPring-8 Center, Sayo,
Hyogo 679-5198, Japan} \email[]{samuel.tardif@spring8.or.jp}
\affiliation{Univ. Grenoble Alpes, Inst NEEL, F-38042 Grenoble, France} 
\affiliation{CNRS, Inst NEEL, F-38042 Grenoble, France} 
\affiliation{CEA, INAC, SP2M
Grenoble 38054 France}

\author{Andrey~Titov}
\affiliation{Univ. Grenoble Alpes, Inst NEEL, F-38042 Grenoble, France} 
\affiliation{CNRS, Inst NEEL, F-38042 Grenoble, France}
\affiliation{CEA, INAC, SP2M Grenoble 38054 France}
\affiliation{A. M. Prokhorov General Physics Institute, Russian Academy of Sciences, 38 Vavilov street, Moscow 119991 Russia}

\author{Emmanuel~Arras}
\affiliation{CEA, INAC, SP2M Grenoble 38054 France}

\author{Ivetta~Slipukhina}
\affiliation{CEA, INAC, SP2M Grenoble 38054 France}

\author{El-K\'{e}bir~Hlil}
\affiliation{Univ. Grenoble Alpes, Inst NEEL, F-38042 Grenoble, France} 
\affiliation{CNRS, Inst NEEL, F-38042 Grenoble, France}

\author{Salia~Cherifi}
\affiliation{IPCMS, CNRS-UdS, 23 rue du Loess, F-67034 Strasbourg, France}

\author{Yves~Joly}
\affiliation{Univ. Grenoble Alpes, Inst NEEL, F-38042 Grenoble, France} 
\affiliation{CNRS, Inst NEEL, F-38042 Grenoble, France}

\author{Matthieu~Jamet}
\affiliation{CEA, INAC, SP2M Grenoble 38054 France}

\author{Andr\'{e}~Barski}
\affiliation{CEA, INAC, SP2M Grenoble 38054 France}

\author{Jo\"{e}l~Cibert}
\affiliation{Univ. Grenoble Alpes, Inst NEEL, F-38042 Grenoble, France} 
\affiliation{CNRS, Inst NEEL, F-38042 Grenoble, France}

\author{Erkin Kulatov}
\affiliation{A. M. Prokhorov General Physics Institute, Russian Academy of Sciences, 38 Vavilov street, Moscow 119991 Russia}

\author{Yurii A. Uspenskii}
\affiliation{P. N. Lebedev Physical Institute, Russian Academy of Sciences, 53 avenue Leninskii, Moscow 119991 Russia}

\author{Pascal~Pochet}
\email[]{pascal.pochet@cea.fr}
\affiliation{CEA, INAC, SP2M Grenoble 38054 France}

\date{\today}

\begin{abstract}
X-ray absorption (XAS) and x-ray magnetic circular dichroism (XMCD) spectra at the L$_{2,3}$ edges of Mn in (Ge,Mn) compounds have been measured and are compared to the results of first principles calculation. 
Early \textit{ab initio}
studies show that the Density Functional Theory (DFT) can very well describe the valence band electronic properties but fails to 
reproduce a characteristic change of sign in the L$_{3}$ XMCD
spectrum of Mn in Ge$_3$Mn$_5$, which is observed in experiments. In
this work we demonstrate that this disagreement is partially related
to an underestimation of the exchange splitting of Mn 2$p$ core
states within the local density approximation. It is shown that the
change in sign experimentally observed is reproduced if the
exchange splitting is accurately calculated within the Hartree-Fock
approximation, while the final states can be still described by the
DFT. This approach is further used to calculate the XMCD in
different (Ge,Mn) compounds. It demonstrates that the agreement between experimental and theoretical spectra can be improved by combining state of the art calculations for the core and valence states respectively.
\end{abstract}


\maketitle

\section{INTRODUCTION}

The development of spintronics has emphasized the need for novel
materials exhibiting a strong electric-magnetic interplay, as it
would permit the design of new devices achieving an electric control
of the magnetic properties, strongly spin-polarized currents and
magneto-transport effects, or magneto-optical functions.

One class of such materials is that of diluted magnetic
semiconductors (DMSs) presenting carrier induced
ferromagnetism.\cite{Ohno98} Magnetic impurities introduced in II-VI
or III-V semiconductors exhibit a strong coupling to the carriers of
the semiconductor, giving rise to giant magneto-optical (giant Zeman
effect) and magneto-transport properties. If electrically doped,
DMSs also feature ferromagnetic interactions \cite{Ohno98} that
depend on the carrier density in such a way\cite{Dietl00} that the
Curie temperature\cite{Chiba03} and the magnetic
anisotropy\cite{Chiba08} can be controlled in field effect devices.
It was rapidly recognized that in the quest for such materials,
methods complementing magnetic studies were needed: the most widely
used are magneto-optical spectroscopy (giant Zeeman effect and
magnetic circular dichroism at the bandgap),\cite{Ando03}
magneto-transport (magnetoresistance, anomalous Hall
effect),\cite{Jungwirth02} and x-ray spectroscopy (x-ray absorption
spectroscopy, XAS, and x-ray magnetic circular dichroism, XMCD, at
both K and L edges).

XAS and XMCD are two well established techniques for the study of
the electronic and magnetic properties of materials. The existence
of dichroism sum rules, which have been derived theoretically
\cite{Thole, Carra} and then successfully applied to XMCD
experimental spectra,\cite{Chen} makes it straightforward to extract
quantitative information on the local spin and orbital magnetic
moments, in particular in the case of the L$_{2,3}$ absorption edges
of the transition metals. There are several efficient ways of calculating the
theoretical XAS and XMCD spectra  \textit{e.g.} the multiplet approach for localized systems or the configuration interaction (CI) approach for metal systems, where the hybridization between the transition metal $d$ states and the surrounding delocalized states is taken into account as a superposition of different $d$ configuration (\textit{i.e.} different multiplet structures) \cite{Jo91}. 
Additionally the density functional theory (DFT) can accurately describe the valence band electronic properties for both ionic and metallic systems and therefore it can also be used to calculate the theoretical XAS and XMCD spectra.
Both the sum rules and DFT calculations at the L$_{2,3}$ absorption edges work well for heavier 3\textit{d} transition metals (\textit{i.e.} Fe, Co), however they are more difficult to apply to lighter ones. \cite{Piamonteze09,Wende07}
A discrepancy as high as 50 to 80\% or even a wrong sign has been reported
between the value of the magnetic spin moment in 3$d^{4}$ Mn$^{3+}$ deduced from the
application of the sum rule, and its expectation
value (Note that for a 3\textit{d}$^{5}$ system the error is much reduced to a value between 68\% and 74\% with no wrong sign).\cite{Piamonteze09}   The difficulty in such an approach is that
interaction between the photocreated 2$p$-core hole and the
3$d$-electrons of Mn modifies considerably the shape of L$_{2,3}$
spectra, leading to multiplet effects.\cite{deGrootBook, Sangaletti10} These
effects are not always caught by DFT, and a disagreement between
experimental and theoretical L$_{2,3}$ spectra is often found.\cite{Sangaletti10} 

As a result, it would be highly desirable to improve the DFT-based XAS and XMCD calculations for lighter transition metals.
In particular, Mn is the transition metal impurity which is the most
widely used to make DMSs. In II-VI semiconductors such as selenides
and tellurides, all compositions up to 100\% Mn can be grown by
molecular beam epitaxy. The most studied DMS with carrier induced
ferromagnetism is (Ga,Mn)As where Mn substitutes Ga.  Values of the
Curie temperature have been improved but stay lower than 200~K. As
this is too low for practical applications, and in order to ensure a
good compatibility with silicon technology, some effort has been
directed towards introducing Mn into germanium. Recent reports have
shown significantly higher Curie temperatures, but also that the
distribution of magnetic impurity ions is
inhomogeneous.\cite{Chen02,Moreno02,Li05} Such nanostructures,
\textit{e.g.} inclusions, contain a locally high concentration of Mn
ions. Thus the observed high temperature ferromagnetism can be
explained by a stronger exchange interaction between Mn ions, which
are separated by shorter distances in the inclusion. Nevertheless,
interesting magneto-transport properties have been reported. This
class of hybrid systems \cite{Dietl10} which exhibit high values of
the Curie temperature and strong magneto-transport and/or
magnetooptical properties, comprises also, \emph{e.g.},
(Zn,Cr)Te,\cite{Kuroda07} (Ga,Mn)N and (Ga,Fe)N\cite{Bonanni08}, and
MnAs in GaAs,\cite{Hai09} to cite but a few examples.

Our main interest here is driven by understanding the electronic,
magnetic and structural properties of self-assembled ferromagnetic
(Ge,Mn) nanocolumns, a system in which high $T_{C}$ ($>$400~K) have
been reported.\cite{Jamet06} The structural properties of the
nanocolumns and surrounding Ge matrix have already been studied
experimentally\cite{TardifAPL10,TardifPRB10,Rovezzi08,Devillers07}
and theoretically,\cite{Arras10,Arras12} yet an experimental
confirmation of the crystal phase in the nanocolumns is still to be
obtained. Note that similarly high values of the Curie temperature
have been found in other (Ge,Mn) samples with a different
morphology.\cite{Xiu}

Turning to XAS and XMCD at the L$_{2,3}$ edges, the XAS line is
generally broader than the XMCD one. In intermetallic compounds, it
does not show the multiplet structure\cite{Kimura97} which is
usually present in oxides.\cite{TardifAPL10} It is therefore
difficult to use XAS to investigate the local atomic structure
around absorbers.\cite{Sangaletti10} XMCD spectra have a more
developed structure and in this work they are used to study local
atomic structure around Mn atoms in different compounds. However, sum
rules are difficult to use in the present case, first because the
two components L$_2$ and L$_3$ overlap, and second because we deal
with a non-homogeneous system. 

The XMCD spectra measured in both (Ge,Mn) nanocolumns and in the metallic Ge$_3$Mn$_5$
alloy\cite{Hirai04,DePadova08, Sangaletti10} (see Fig.~\ref{fig1}
below) show a clear positive bump between the L$_3$ and L$_2$ lines which could not be reproduced by DFT calculations so far. \cite{Picozzi04,Sangaletti10} 
This disagreement partially comes from an underestimation of the
exchange splitting of Mn 2$p$-states within the local density
approximation (LDA). Typically, the XMCD is calculated by taking into account the spin-orbit coupling (SOC) in the initial (core) states and the exchange interaction in the final (valence) states \cite{Erskine75}. However, it is known that while it does not affect the absorption spectra, taking into account the exchange interaction for the core states in transition metals can have a remarkable impact on the calculated XMCD spectra, mostly around the L$_3$ edge.\cite{Ebert96, Ebert96b} On the opposite, it was shown that taking into account the SOC in the valence states has only a very limited impact on the calculated spectra.\cite{Ebert96, Ebert96b} 

In this paper we start from the assumption that the exchange interaction between 2$p$- and 3$d$-electrons of Mn is weakly screened by valence electrons and we propose to evaluate the splitting of the core states within the Hartree-Fock (HF) approximation, whereas the valence states in metals are still calculated within DFT. We apply this heuristic approach to the calculation of the XAS-XMCD spectra of Mn and test it in the specific case of the ferromagnetic semiconductor (Ge,Mn) and related systems. This is done in a fashion that may be further applied to other systems.

\section{METHODS}
\subsection{Experimental}
\begin{figure}
\includegraphics[height=8.0cm, width=7.0cm, angle=270]{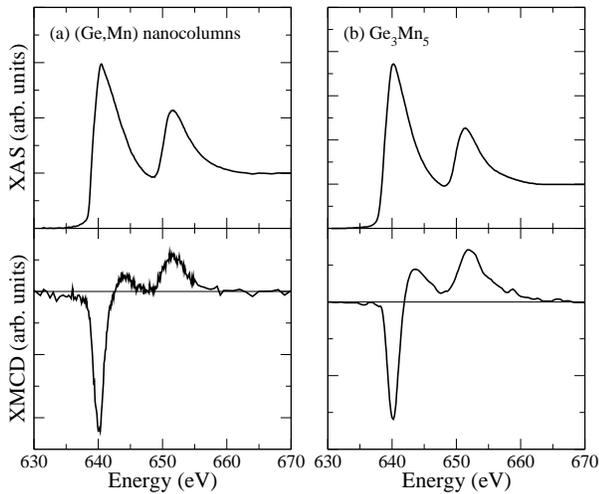}
\caption{\label{fig1} (a) Experimental XAS and XMCD spectra in the
(Ge,Mn) nanocolumns measured at a temperature of 5~K and in a 5~T
magnetic field and (b) in Ge$_{3}$Mn$_{5}$ at 80~K and in remanence
(from Ref.~\onlinecite{DePadova08}).}
\end{figure}

Experimental XAS and XMCD spectra measurements in (Ge,Mn) nanocolumn samples were carried out on 80-nm-thick thin films obtained by low-temperature molecular beam epitaxy. Details on
the sample growth can be found in Ref.~ \onlinecite{Devillers07}.
The (Ge,Mn) samples were efficiently protected against oxidation by
\textit{in situ} deposition of a 3~nm-thick amorphous Si layer.
Measurements were carried out at beamline UE46-PGM at the Helmholtz
Center Berlin using the total electron yield
method.\cite{TardifAPL10} A magnetic field of 5~T has been applied
in the plane of incidence of the x-rays to align the magnetization
along the light propagation. The sample temperature was 5~K.
Experimental XAS-XMCD spectra in the (Ge,Mn) nanocolumns system are
shown in Fig.~\ref{fig1}. No saturation effects\cite{Nakajima99} could be evidenced from measurements at various incidence angles ($30^{\circ}$, $60^{\circ}$ and $90^{\circ}$).

We also considered XAS-XMCD spectra of the ferromagnetic metal
Ge$_{3}$Mn$_{5}$, as reported in the literature both for bulk single
crystals \cite{Hirai04} and for thin films, \cite{DePadova08,
Sangaletti10} see Fig.~\ref{fig1}. They appear to be quite similar
to those of the (Ge,Mn) nanocolumns.

\subsection{Model}
We now turn to the calculation of the theoretical XAS/XMCD spectra. To obtain the spectra, a three step calculation was performed: first the crystal potential was accurately calculated \textit{ab initio}. \textit{Ab initio} methods are based on an explicit band structure calculation, where hybridization of Mn atoms with neighbor atoms and related charge redistributions are taken into account. This method is therefore more suitable to predict XMCD spectra in compounds with different local atomic structure. Then the exchange interaction in the core states was calculated within the Hartree-Fock approximation which allows for an accurate description of the core states. Finally the theoretical XAS/XMCD spectra were calculated within the multiple scattering approach. Each step is described in details hereafter.

\subsubsection{Crystal potential}
The crystal potential was obtained from a self-consistent solution of the Kohn-Sham equations within the general gradient approximation (GGA)\cite{GGA} using the full-potential WIEN2k code.\cite{wien2k}
The spherical part of the self-consistent crystal potential in each MT-sphere was retained for the XAS calculations. 
Then the Schr\"{o}dinger (or Dirac) equation was solved in each MT-sphere.

The experimental lattice parameter of bulk Ge $a$=5.66~\r{A} was used
in all compounds, except in Ge$_2$Mn(C16),\cite{Arras10} and
Ge$_3$Mn$_5$. For the Ge$_2$Mn(C16) compound $a$=5.95~\r{A} and
$c$=5.03~\r{A} parameters were obtained from a structural relaxation
calculation,\cite{Arras10} while experimental values\cite{Forsyth90}
$a$=7.184~\r{A} and $c$=5.053~\r{A} were used for Ge$_3$Mn$_5$. In
all the structures atomic positions were relaxed, except for
Ge$_3$Mn$_5$, where experimental atomic positions were taken from
Ref.~\onlinecite{Forsyth90}. Note that a previous calculation shows
that a relaxation of internal atomic positions and lattice
parameters within GGA does not change the experimental
parameters.\cite{Picozzi04,Slipukhina09,Arras11}

In most cases the Kohn-Sham eigenvalues and eigenfunctions,
calculated in metals, are rather close to those obtained from the
solution of the quasiparticle equation\cite{Hedin70}:

\begin{eqnarray}
\nonumber
\label{eq3}
\left [ - {\frac{1}{2}} \nabla^{2} + V_{ext} (\mathbf{r}) + V_{H} (\mathbf{r}) \right ] \phi_{n\mathbf{k}} (\mathbf{r}) \\
+ \int \Sigma (\mathbf{r},\mathbf{r}',E_{n\mathbf{k}}) \phi_{n\mathbf{k}} (\mathbf{r}') d\mathbf{r}' = E_{n\mathbf{k}} \phi_{n\mathbf{k}} (\mathbf{r}).
\end{eqnarray}

Therefore it is assumed that final state eigenfunctions $\phi_f$ and
eigenvalues $\varepsilon_f$ are well described by DFT in metallic
compounds. Meanwhile the DFT eigenvalues of the initial states can
be improved, as shown in the next step.

\subsubsection{Core levels exchange interaction}
The initial states in x-ray absorption spectroscopy at the L$_{2,3}$
edges are the $2p$ core levels of Mn. They are split by the
spin-orbit and $2p$-$3d$ exchange interactions, described by the
$\xi$ and $H_{xc}$ parameters correspondingly. When $H_{xc}$=0, the
$2p$ core levels are split into a $p_{3/2}$ and a $p_{1/2}$ states,
separated by a $3\xi/2$ energy interval.\cite{Laan95} The spin-orbit
splitting in a spherical atomic potential is given by the term:

\begin{eqnarray}
\nonumber
\label{eq5}
\zeta(r) \mathbf{l} \cdot \mathbf{s},
\end{eqnarray}

where $\zeta(r) \sim \frac{1}{r} \frac{dV}{dr}$ and $V$ is the
atomic potential. This term is large for core orbitals, since they
are localized near the atomic nucleus. The experimental spin-orbit
splitting of the 2$p_{3/2}$ and 2$p_{1/2}$ orbitals is $3\xi/2$=
10.5~eV in Mn ($\xi$=7.0~eV).\cite{Ishiwata02}

The exchange splitting $H_{xc}$ of 2$p_{j=3/2}^{m_j=3/2}$ and
2$p_{j=3/2}^{m_j=-3/2}$ is smaller, $H_{xc}$/$\xi\ll$1. This term
arises from the exchange interaction between 2$p$- and
3$d$-electrons of the Mn atom. 
Considering that the exchange coupling acts only on the spin of the core states, the interaction Hamiltonian can be written as $gSH_{xc}$, where $g$ is the electron gyromagnetic factor and $S$ is the spin operator. 
As a result, the 2$p$-level with $j$=3/2 is split into four successive sublevels ($m_j$=-3/2, -1/2, 1/2, 3/2) and the 2$p$-level with $j$=1/2 is split into two sublevels ($m_j$=1/2, -1/2). The energy splitting between two consecutive sublevels is given in units of $gH_{xc}$ by:

\begin{eqnarray}
\label{eq2_0}
\frac{1}{2}+\frac{s(s+1)-l(l+1)}{2j(j+1)}
\end{eqnarray}

where $s = 1/2$ and $l = 1$ for 2$p$ electronic states. 
The energy separation of each sublevel is then $-gH_{xc}/3$ at the $j=3/2$ level and $gH_{xc}/3$ at the $j=1/2$ level. 
Therefore, state $m_j$=3/2 is the lowest in energy at the 2$p_{3/2}$ level and conversely, state $m_j$=-1/2 is the lowest in energy at the 2$p_{1/2}$ level.
Furthermore, the value of $H_{xc}$ can be evaluated as a difference between eigenvalues of the $2p$-core sublevels and approximating $g$ by 2:

\begin{eqnarray}
\label{eq2}
2H_{xc} = \epsilon_{j=3/2}^{m=-3/2} - \epsilon_{j=3/2}^{m=3/2}.
\end{eqnarray}

If the exchange interaction between the 2$p$- and 3$d$-electrons of a Mn atom is weakly screened by valence electrons, a good estimate of the value of the exchange splitting $H_{xc}$ can be obtained within the Hartree-Fock approximation, from a calculation of the exchange term in the self-energy:

\begin{eqnarray}
\nonumber
\label{eq4}
2H_{xc} = &\left\langle\phi_{j=3/2}^{m=-3/2}|Gv|\phi_{j=3/2}^{m=-3/2}\right\rangle \\
 &-\left\langle\phi_{j=3/2}^{m=+3/2}|Gv|\phi_{j=3/2}^{m=+3/2}\right\rangle,
\end{eqnarray}

where the Green's function is composed by the $3d$-orbitals of the
Mn atom, $v$ is the bare Coulomb potential, and $\phi_j^m$ are
$2p$-core orbitals of the same Mn atom, calculated by the DFT
method. It is expected that the contribution of other terms of the
quasiparticle equation (\ref{eq3}) is smaller since the radial parts
of $\phi_j^m$ orbitals with $j$=3/2,$m_j$=-3/2 and
$j$=3/2,$m_j$=+3/2 are almost identical.

In most cases, the spin-orbit interaction between valence electrons
was neglected in our calculation (\emph{i.e.}, the Schr\"{o}dinger
equation was solved in each MT-sphere) whereas the spin-orbit
splitting of the $2p$ core states was always taken into account. In
order to evaluate the effect of spin-orbit interaction between
valence electrons on absorption spectra, fully relativistic
calculations of final states in Eq.~(\ref{eq1}) were also performed.

\subsubsection{Absorption and magnetic dichroism spectra}
Finally, XAS spectra of (Ge,Mn) compounds have been computed from first principles using the FDMNES code.\cite{FDMNES} 
In the dipole approximation, the x-ray absorption cross-section is given by:

\begin{eqnarray}
\label{eq1}
\sigma (\omega) =  4 \pi^{2} \alpha \hbar \omega \sum_{i,f} \left| <\phi_{f} |\hat{\epsilon}| \phi_{i}> \right|^{2} && \nonumber \\
\times \delta_{\Gamma} (\hbar \omega - \varepsilon_f + \varepsilon_i),&&
\end{eqnarray}

where $\alpha$ is the fine structure constant, $\phi_i$,
$\varepsilon_i$, and $\phi_f$, $\varepsilon_f$ are eigenfunctions
and eigenvalues of the initial and final states correspondingly,
$\delta_{\Gamma}$ is a Lorentzian curve of width $\Gamma$ determined
by the core-hole lifetime, and $\hat{\epsilon_{r}}$ is the photon polarization.
The absorption cross-section is calculated for the three orthogonal
directions of light propagation (\textit{x},\textit{y},\textit{z}),
and the direction of the magnetization along the light propagation.
Then a mean value of the cross-section is evaluated.

The electron final states in the cluster were calculated using the
multiple-scattering approach and within the muffin-tin (MT)
approximation.\cite{Natoli86,Zabinsky95} The calculation was
performed in a periodic crystal potential, but scattering paths were
considered in a cluster with a diameter of 12~\r{A}.

The XAS and XMCD signals were then evaluated as
($\sigma^{+}$+$\sigma^{-}$)/2 and ($\sigma^{+}$-$\sigma^{-}$),
respectively, where $\sigma^{+}$ and $\sigma^{-}$ are absorption
cross-sections for the two circular polarizations of x-ray
radiation.

Calculated XAS and XMCD spectrum at the L$_{2,3}$-edges of Mn were convoluted with a Lorentzian function to account for the core-hole lifetime mentioned above (for the sake simplicity the same value $0.32~eV$ was used for both the L$_{2}$ and L$_{3}$-edges). A complementary convolution with a Gaussian function to account for the experimental resolution and scaling of the spectra was performed to compare the calculated spectra with the experimental ones. Typical broadening parameters $\sigma=1.0~eV$ and $\sigma=0.5~eV$ (where $\sigma$ is the standard deviation of the Gaussian function) were used\cite{Picozzi04}. The smaller value was used for the XMCD spectra so as not to completely wash out the structure of the peaks that will be discussed hereafter.

\section{RESULTS}

\subsection{Exchange splitting of 2\textit{p}-core levels}

\begin{figure}
\includegraphics[height=8.0cm, width=7.0cm, angle=270]{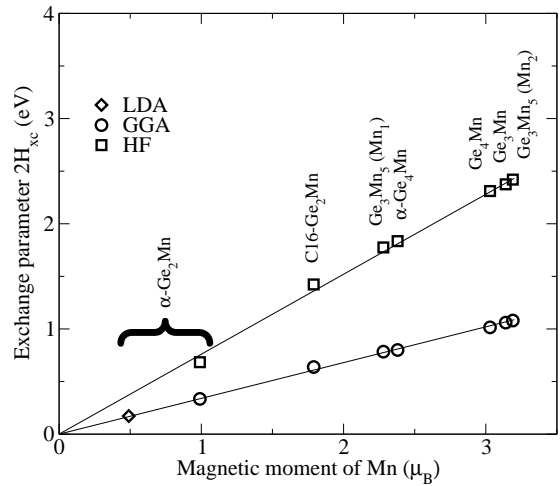}
\caption{\label{fig2} Exchange splitting of Mn 2$p$-states in
different compounds (see text), calculated within the GGA (circles)
and the HF approximation (squares). Numerical values are given in
Tab.~\ref{tab1}. The diamond symbol corresponds to the
$\alpha$-Ge$_{2}$Mn compound calculated within the LDA.}
\end{figure}

The exchange splitting $H_{xc}$ in different compounds (these
compounds are described in more details in the next sections) was
calculated within the GGA and the HF approximation. Calculation
parameters and corresponding exchange splitting values are listed in
Table~\ref{tab1}. $H_{xc}$ is plotted as a function of Mn spin
moment in Fig.~\ref{fig2}.

The dependence of the exchange splitting $H_{xc}$ on Mn spin moment
is almost linear, and actually small deviations from the linearity
can be related to different radii of MT-spheres which were used to
compute Mn spin moments in the different compounds.

The value of $H_{xc}$ is greatly underestimated within LDA and GGA.
This shows the inadequacy of both LDA and GGA to describe correctly
such an effect. Note that within the LDA (illustrated only for the
$\alpha$-Ge$_{2}$Mn compound in Fig.~\ref{fig2} for clarity), not
only the exchange splitting but also the magnetic moment on the Mn
atom are underestimated.

\begin{table*}[ht]\footnotesize
\caption{Values of structural and magnetic parameters obtained from
the WIEN2K calculation: R$_{MT}$ is the radius of a
muffin-tin sphere around the atom; M$_{local}$ is the magnetic
spin moment of the atom, calculated by integration of the spin
density over the sphere; M$_{TOT}$ is the magnetic spin
moment per formula unit. The last parameter, $2H_{xc}$, is the splitting
parameter introduced in Eq.~\ref{eq4}, calculated in the HF approximation. For each compound, the
number of atoms of each sort is indicated by the multiplication
factor.}

\centering
\resizebox{18cm}{!} {
\begin{tabular}{|c|c|c|c|c|c|c|c|c|c|c|c|c|c|c|c|c|}

\hline\hline

         & \multicolumn{3}{c|}{Ge$_3$Mn$_5$}  & \multicolumn{2}{c|}{Ge$_2$Mn-$\alpha$} & \multicolumn{2}{c|}{Ge$_4$Mn-$\alpha$} & \multicolumn{2}{c|}{Ge$_2$Mn-C16} & \multicolumn{3}{c|}{Ge$_3$Mn} & \multicolumn{4}{c|}{Ge$_4$Mn} \\
\hline
         & 2$\times$Mn$_1$ & 3$\times$Mn$_2$ & 3$\times$Ge$_1$ & 1$\times$Mn$_1$ & 2$\times$Ge$_1$ & 1$\times$Mn$_1$ & 4$\times$Ge$_1$ & 1$\times$Mn$_1$ & 2$\times$Ge$_1$ & 1$\times$Mn$_1$ & 1$\times$Ge$_1$ & 2$\times$Ge$_2$ & 1$\times$Mn$_1$ & 1$\times$Ge$_1$ & 1$\times$Ge$_2$ & 2$\times$Ge$_3$ \\
\hline
R$_{MT}$, \r{A} & 1.241 & 1.241 & 1.241 & 1.254 & 1.254 & 1.281 & 1.281 & 1.201 & 1.249 & 1.185 & 1.138 & 1.138 & 1.185 & 1.138 & 1.138 & 1.138 \\
\hline
M$_{local}$, $\mu_B$ & 2.28 & 3.19 & -0.15 & 1.00 & -0.02 & 2.38 & -0.04 & 1.79 & -0.08 & 3.14 & 0.03 & -0.13 & 3.03 & 0.00 & 0.02 & -0.01 \\
\hline
M$_{TOT}$, $\mu_B$ & \multicolumn{3}{c|}{13.73}  & \multicolumn{2}{c|}{0.96} & \multicolumn{2}{c|}{2.19} & \multicolumn{2}{c|}{1.58} & \multicolumn{3}{c|}{2.98} & \multicolumn{4}{c|}{3.16} \\
\hline
$2H_{xc}$, eV & 1.775 & 2.420 &  & 0.684 &  & 1.835 &  & 1.423 &  & 2.375 &  &  & 2.311 &  &  &  \\
\hline\hline

\end{tabular}
}
\label{tab1}
\end{table*}

\subsection{Ge$_3$Mn$_5$, a test material}
As mentioned earlier, the intermetallic compound Ge$_3$Mn$_5$
displays experimental XAS-XMCD spectra similar to those observed in
the (Ge,Mn) nanocolums (Fig.~\ref{fig1}). Accordingly, we start our
detailed study by considering this well-known system, which is
available in bulk form and as thin epitaxial layers. It has a
hexagonal crystal structure with a space group P6$_{3}$/$mcm$ and
lattice parameters $a$=7.184~\r{A} and
$c$=5.053~\r{A}.\cite{Forsyth90} The primitive cell contains three
inequivalent atoms, Mn$_1$, Mn$_2$ and Ge, in
positions:\cite{Picozzi04}

4(d) Mn$_1$:$\pm$($\frac{1}{3}$,$\frac{2}{3}$,0),$\pm$($\frac{2}{3}$,$\frac{1}{3}$,$\frac{1}{2}$)

6(g) Mn$_2$:$\pm$($x$,0,$\frac{1}{4}$),$\pm$(0,$x$,$\frac{1}{4}$),$\pm$($-x$,$-x$,$\frac{1}{4}$), $x$=0.2397

6(g) Ge:$\pm$($x$,0,$\frac{1}{4}$), $\pm$(0,$x$,$\frac{1}{4}$), $\pm$($-x$,$-x$,$\frac{1}{4}$), $x$=0.6030

The experimental magnetic spin moments of Mn$_1$ and Mn$_2$ atoms
are different: 1.96(3)~$\mu_B$ and 3.23(2)~$\mu_B$ correspondingly
(calculated values are listed in Table~\ref{tab1}). This difference
was attributed to different atomic structures around the Mn
atoms.\cite{Forsyth90} The local atomic structure of Mn atoms in
Ge$_3$Mn$_5$, as given in Ref.~\onlinecite{Forsyth90}, is the
following: (\textit{i}) Mn$_1$ neighbors are two Mn$_1$, six Ge and
six Mn$_2$ at distances 2.522~\r{A}, 2.534~\r{A} and 3.059~\r{A}
respectively; (\textit{ii}) Mn$_2$ neighbors are two Mn$_2$, four
Mn$_2$ and four Mn$_1$ at distances 2.976~\r{A}, 3.051~\r{A} and
3.059~\r{A} respectively. The two Mn atoms nearest-neighbors to
Mn$_1$ atoms are at a short distance (2.522~\r{A}) and it was
suggested that the interaction with these neighbors would lead to a
reduction of the magnetic spin moment of Mn$_1$ atoms. The
calculated total spin moment per Mn atom is 2.75~$\mu_B$, in
agreement with previous calculations.\cite{Picozzi04, Stroppa06, Stroppa07, Slipukhina09}
This value is slightly higher than the experimental one,
\textit{i.e.} 2.60~$\mu_B$.\cite{Forsyth90} It was shown that a good
agreement with experiment is found within GGA and including the
spin-orbit interaction into the calculation.\cite{Picozzi04}

\begin{figure}
\includegraphics[height=8.0cm, width=7.0cm, angle=270]{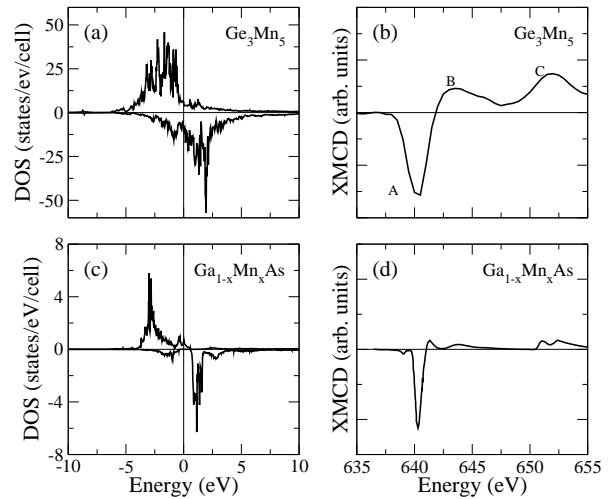}
\caption{\label{fig3} 3$d$-states of Mn in (a) Ge$_3$Mn$_5$ and (c)
Ga$_{1-x}$Mn$_x$As ($x$=0.125). Experimental XMCD spectra at the
L$_{2,3}$ edges of Mn in (b) Ge$_3$Mn$_5$ (from
Ref.~\onlinecite{DePadova08}) and (d) Ga$_{1-x}$Mn$_x$As ($x$=0.084,
from Ref.~\onlinecite{Edmonds06}).}
\end{figure}

The partial density of 3$d$-states of Mn in Ge$_3$Mn$_5$ is shown in
Fig.~\ref{fig3}. It is interesting to compare it with the Mn 3$d$
DOS in a diluted magnetic semiconductor Ga$_{1-x}$Mn$_x$As
($x$=0.125), where Mn atoms substitute Ga. The density of states at
the Fermi level in Ge$_3$Mn$_5$ is significant in both spin channels
and this compound has a metallic conductivity, as it was found in
previous calculations.\cite{Picozzi04}

Because of a strong interaction between Mn atoms, the 3$d$-bands in Ge$_3$Mn$_5$ are broad. The L$_{2,3}$ spectra reflect electron transitions from the narrow $2p$- into the broad valence $3d$-bands of Mn, hence the lineshapes in XMCD spectra at the L$_{2,3}$ edges of Mn in intermetallic compounds is determined by the width of the valence bands.

It can be seen from
Fig.~\ref{fig3}, that the L$_2$ and L$_3$ edges are narrow in
Ga$_{1-x}$Mn$_x$As ($x$=0.125), in agreement with the narrow bands
calculated for this DMS.

\begin{figure}
\includegraphics[height=8.0cm, width=7.0cm, angle=270]{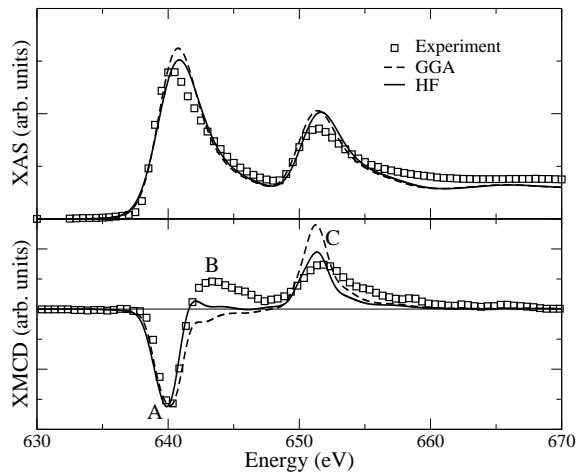}
\caption{\label{fig4} Calculated (a) XAS and (b) XMCD spectra of
Ge$_3$Mn$_5$ at the L$_{2,3}$-edges of Mn. The exchange splitting of
the Mn  2$p$ core levels was evaluated within GGA (dash line) and HF
approximation (solid line). Experimental data were taken from
Ref.~\onlinecite{DePadova08}.}
\end{figure}

The resulting XAS and XMCD spectrum, calculated within GGA and HF
approximation, are shown in Fig.~\ref{fig4}. XAS obtained within
both approximations are similar and contain L$_2$ and L$_3$ edges
without visible multiplet structure. A large value of the
convolution parameter $\sigma=1~eV$ in Gaussian functions allows us
to reproduce the broad absorption lines found in experimental
L$_{2,3}$ spectra of Mn.

The XMCD spectrum contains a fine structure, which can be used to
distinguish compounds with different atomic structure around Mn
atoms. When all the 3$d$-states with spin up are occupied, the XMCD
spectrum shows a negative A peak and a positive C peak.\cite{Laan99}
In solids, the 3$d$ states of Mn are partially occupied and the XMCD
spectrum has a more complicated structure: the main absorption line
A in the L$_3$ edges is followed by a positive peak B
(Fig.~\ref{fig3}). This feature of the absorption line is not
reproduced by the DFT-GGA calculation,\cite{Picozzi04}
however a description of Mn 2$p$-3$d$ exchange interaction within
the HF approximation allows us to obtain a qualitative agreement
with experiment. In particular, line B is reproduced, although its
intensity is smaller than in experiment (Fig.~\ref{fig4}). The
intensity ratio between the two absorption edges L$_2$ and L$_3$ is
improved as well.

\begin{figure}
\includegraphics[height=8.0cm, width=7.0cm, angle=270]{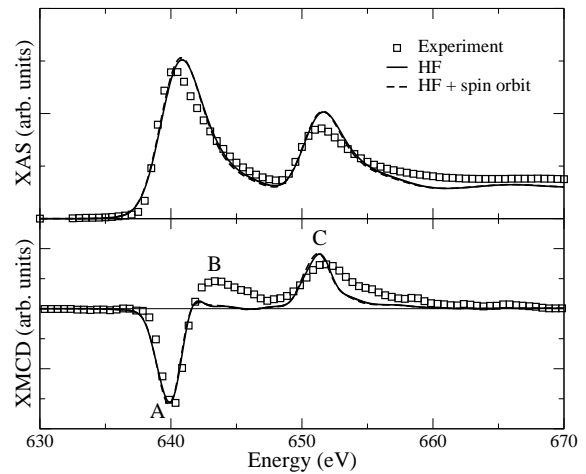}
\caption{\label{fig5} Calculated (a) XAS and (b) XMCD spectra of
Ge$_3$Mn$_5$ at the L$_{2,3}$-edges of Mn with (dashed line) and
without (solid line) spin-orbit interaction. Both spectra are almost
superimposed, showing the small influence of the spin-orbit
interaction in this metallic compound. The exchange splitting of Mn
 2$p$ core levels was evaluated within HF approximation.}
\end{figure}

The influence of spin-orbit interaction on the absorption spectrum
was studied by solving the Dirac equation for the initial and final
states. Then the exchange splitting of Mn 2$p$ sublevels was
corrected and the absorption spectrum was calculated according to
Eq.~(\ref{eq1}). The influence of spin-orbit interaction on XAS and
XMCD spectrum was found to be small (Fig.~\ref{fig5}), and
essentially the same spectrum is obtained by solving the
Shr\"{o}dinger equation with spin polarization in MT-spheres.

\subsection{The $\alpha$-Ge$_{2}$Mn structure}
Another structure was initially proposed by Takizawa \textit{et
al.}\cite{Takizawa} for Ge$_{4}$Mn, and further discussed by Arras
\textit{et al.}\cite{Arras12} for $\alpha$-Ge$_{2}$Mn as a likely
candidate for the crystalline structure of the nanocolumns in
(Ge,Mn) layers. This structure is derived from cubic Ge, in which
the presence of interstitial Mn atoms lowers the formation energy.
The formation energy of the $\alpha$-Ge$_{2}$Mn was also found to be
lower than that of Ge with the usual diamond structure containing
the same amount of substitutional and interstitial Mn. This phase is
all the more favorable when the concentration of substitutional or
interstitial Mn increases.\cite{Arras12}

\begin{figure}
\includegraphics[height=8.0cm, width=7.0cm, angle=270]{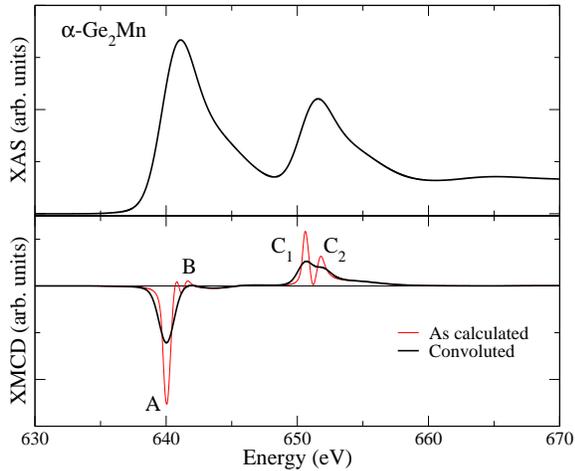}
\caption{\label{fig6} (color online) Calculated (a) XAS and (b) XMCD spectra of
$\alpha$-Ge$_{2}$Mn at the L$_{2,3}$-edges of Mn. The calculated XMCD
spectrum is shown by a thin red (gray) line. The same spectrum after
convolution with the Lorentzian and Gaussian functions is shown by a
thick black line. The exchange splitting of Mn  2$p$ core levels was
evaluated within the HF approximation.}
\end{figure}

\begin{figure}
\includegraphics[height=8.0cm, width=7.0cm, angle=270]{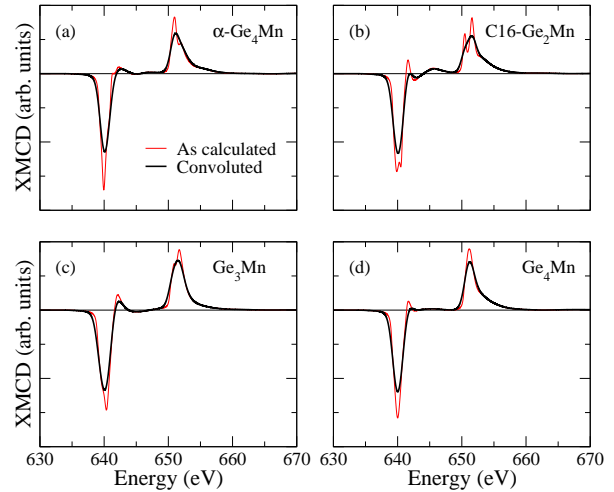}
\caption{\label{fig7} (color online) Calculated XMCD spectra at the L$_{2,3}$-edges
of Mn in different (Ge,Mn) compounds (a) $\alpha$-Ge$_{4}$Mn, (b)
C16-Ge$_{4}$Mn, (c) Ge$_3$Mn with substitutional Mn in Ge with
diamond structure and (d) Ge$_4$Mn with interstitial Mn in Ge with
the diamond structure. The calculated XMCD spectrum is shown by a
thin red (gray) line. The same spectrum after convolution with the
Lorentzian and Gaussian functions is shown by a thick black line.
The exchange splitting of Mn  2$p$ core levels was evaluated within
the HF approximation.}
\end{figure}

The XAS and XMCD spectra of $\alpha$-Ge$_{2}$Mn were calculated and
convoluted with Lorentzian and Gaussian functions in the same manner
as it was done for Ge$_3$Mn$_5$. The resulting spectra are shown in
Fig.~\ref{fig6}. As in the case of Ge$_3$Mn$_5$, it is predicted
that XAS of $\alpha$-Ge$_{2}$Mn has no fine structure so that it can
hardly be used to study the local atomic environment of Mn.

In contrast, the XMCD spectrum has a more detailed structure
(Fig.~\ref{fig6}), and for this reason only XMCD spectra will be
further considered. The L$_{2,3}$ XMCD spectrum of Mn contains a
negative line A, followed by a slightly negative broad line B in the
L$_{3}$ edge. The positive line C at the L$_{2}$ edge is not
symmetric: its left side is sharper that the right one. This
asymmetry is due to a splitting of the line C into a more intense
C$_1$ and a less intense C$_2$ lines.

The positive line B, observed in the XMCD spectrum of Ge$_3$Mn$_5$,
is absent in the spectrum of $\alpha$-Ge$_{2}$Mn. The local magnetic
moment of Mn in this structure is low (1$\mu_B$), and the exchange
splitting of Mn 2$p$-states is small. As it was shown in previous
section, the line B is reproduced in calculated spectra of Mn if the
exchange splitting is properly evaluated, \emph{i.e.}, if the
splitting of 2$p$-states underestimated within GGA, is increased
approximately by a factor 2. This suggests that an increase of the
local spin moment of Mn, together with the corresponding increase of
the exchange splitting, would lead to the appearance of the line B
in the XMCD spectrum of $\alpha$-Ge$_{2}$Mn.

As a matter of fact, such an increase of the local spin moment of Mn
is expected if vacancies are created at the Mn positions in
$\alpha$-Ge$_{2}$Mn. In particular, in a $\alpha$-Ge$_{4}$Mn phase
which has the crystal structure of $\alpha$-Ge$_{2}$Mn, but only
50$\%$ of Mn atoms in their original positions, the local spin
moment of Mn is increased to 2.4~$\mu_B$ (Tab.~\ref{tab1}). Then
line B appears in the theoretical XMCD spectrum (Fig.~\ref{fig7}a).

\subsection{The C16-Ge$_{2}$Mn structure}

A Ge$_{2}$Mn phase, related to the $\alpha$-phases, was proposed in
Ref.~\onlinecite{Arras10}. This phase has a tetragonal lattice
structure with a cell volume smaller than the one in bulk Ge. The
calculated spin moment of Mn in C16-Ge$_{2}$Mn is 1.8~$\mu_B$, that
is, larger than the Mn spin moment in $\alpha$-Ge$_{2}$Mn
(1.0~$\mu_B$). This increased value of the Mn spin moment enhances
the exchange splitting of the 2$p$-core states (Fig.~\ref{fig2}).

Different band structures in the $\alpha$- and C16-phases, as well
as different values of the 2$p$-levels exchange splitting, cause
differences in their XMCD spectra. The line B in the XMCD spectrum
of C16-Ge$_{2}$Mn contains a negative part, which is followed by a
positive one (Fig.~\ref{fig7}b). This unusual behavior of the B line
may be used to identify the C16-Ge$_{2}$Mn phase.

\subsection{Ge$_3$Mn and Ge$_4$Mn with diamond structure}

We now consider Ge with the usual diamond structure, where Mn atoms
are located in substitutional or interstitial positions. The Mn
concentrations were taken to be 25$\%$ for substitutional Mn
(\emph{i.e.}, Ge$_3$Mn) and 20$\%$ for interstitial Mn (\emph{i.e.},
Ge$_4$Mn). These compositions are within the range experimentally
observed in different Mn-rich (Ge,Mn) phases,
15$\%$-40$\%$.\cite{Jamet06,Devillers07,Bougeard06}

The calculated XMCD spectra of both substitutional and interstitial
Mn in Ge (Fig.~\ref{fig7}~c,d) are quite close. In  Ge$_3$Mn with
substitutional Mn, the intensity of  line C is larger than in other
phases. In Ge$_4$Mn with interstitial Mn, line B is weaker, in spite
of a large local spin moment on Mn and the related large exchange
splitting of 2$p$-states (Tab.~\ref{tab1}). Also line C is narrower
than in other compounds.

\subsection{Discussion}

We have shown that more accurate simulations of the XMCD spectra can be achieved using the
precise calculation of the core levels splitting. This correct splitting can be obtained from the eigenvalues of initial 2$p$-states, calculated within the HF approximation and under the assumption of a weak screening of the exchange interaction between the 2$p$- and 3$d$- electrons. The validity of the latter is confirmed by our results.
The final valence states can be still described within DFT, since the DFT eigenvalues in metals are rather close to
quasiparticle energies. 

One can note that many-body effects, such as relaxation of the electron system after excitation,\cite{Rehr05} and the mixing of L$_{2}$ and L$_{3}$ edges
due to Coulomb interaction between electrons of Mn,\cite{Teramura96}
are not taken into account in this calculation and may modify the
XMCD spectrum. In addition, the influence of defects on the
electronic state of Mn should be also taken into account when a
comparison to experimental spectra is done, as well as the
disordered structures in the nanocolumn samples, which contain Mn in
the nanocolumns but also in the matrix and at the interface. 
All these factors make it difficult to use XMCD spectra for the
determination of the crystal structures of metallic compounds.

However, we have shown that the more accurate simulation of the XMCD spectra using the
precise calculation of the core levels splitting reveals
details on the XMCD spectra lineshape (\textit{e.g.} the presence or
absence of the positive peak labeled B at about 643 eV) that were up
to now eluded in standard calculations. 

\section{SUMMARY and CONCLUSIONS}

XAS and XMCD spectra at the L$_{2,3}$ edges of Mn in different
(Ge,Mn) compounds were calculated from first principles. Early
calculations show that DFT-based calculations are not able to
reproduce some features in the XMCD spectra of Mn. In particular,
the positive part of the L$_3$ edges of Mn in Ge$_3$Mn$_5$ is absent
in the calculated XMCD spectrum, while it is observed in experiment. In this work we show that this positive part can be reproduced if the core levels splitting is accurately calculated, in agreement with previous results \cite{Ebert96, Ebert96b}.
 The effect of spin-orbit interaction between
valence electrons was also considered and found to be small.

(Ge,Mn) compounds with a large Mn content usually feature a metallic
character. The XAS of such metallic compounds are broad and have no
particularities which can help to identify different compounds. XMCD
spectra have a more detailed structure and their shape depends on
the local spin moment of Mn and the crystal structure. In this work
we have compared XMCD spectra calculated for two energetically stable
phases of (Ge,Mn), with those calculated for substitutional and
interstitial Mn in Ge. The XMCD spectra of all the (Ge,Mn) phases
have a similar structure and shape, but also small peculiar features
which may be used to identify a particular (Ge,Mn) compound.

The method to improve the agreement between theoretical and experimental XMCD spectra that we suggest in this paper can easily be applied to other systems of interest, such as for example transition metal impurities on surfaces where a similar change of sign after the main L$_{3}$ peak has been observed in the experimental XMCD spectra\cite{Gambardella05}, or other homogeneous transition metal compounds.

\begin{acknowledgments}
{Authors acknowledge financial support from the French Research
Agency (project ANR GEMO), the Grenoble Nanosciences Foundation, the Russian Fund for Basic Research (RFBR No 10-02-00698) and the Russian Ministry of Education and Science.
Calculations were performed at the CIMENT supercomputer center via
the NanoSTAR project. We also gratefully acknowledge the assistance
of Detlef Schmitz during experiments at beamline UE46-PGM, Helmholtz
Center, Berlin.}
\end{acknowledgments}

\end{document}